\documentclass[aps,prx,nofootinbib,reprint]{revtex4-1}
\usepackage{amsmath,amssymb}
\usepackage{graphicx}
\usepackage{comment}
\usepackage{booktabs}
\usepackage{accents}
\usepackage{color}
\usepackage{bm}

\definecolor{light-gray}{gray}{0.95}
\newcommand{\shadedbox}[1]{\colorbox{light-gray}{$\displaystyle #1$}}

\newcommand\rr{\mathbf{r}}
\newcommand\pp{\mathbf{p}}
\newcommand\zhat{\mathbf{\hat{z}}}
\newcommand\xhat{\mathbf{\hat{x}}}
\newcommand\yhat{\mathbf{\hat{y}}}
\newcommand\ehatp{\mathbf{\hat{e}_1}}
\newcommand\ehatz{\mathbf{\hat{e}_0}}
\newcommand\ehatm{\mathbf{\hat{e}_{-1}}}
\newcommand\phat{\mathbf{\hat{\pp}}}
\newcommand\rhat{\mathbf{\hat{r}}}
\newcommand\phihat{\mathbf{\hat{\phi}}}

\newcommand\avec{\mathbf{a}}
\newcommand\bvec{\mathbf{b}}

\newcommand\Jrt{\mathbf{J}(\rr,t)}
\newcommand\Jomegagen{\mathbf{J}_\omega(\pp)}
\newcommand\Jomega{\mathring{\mathbf{J}}_\omega(\phat)}
\newcommand\Jomegar{{\mathbf{J}}_\omega(\rr)}
\newcommand\Jomegarhat{{\mathbf{J}}^{r}_\omega(\rr)}
\newcommand\Jomegart{{\mathbf{J}}^{t}_\omega(\rr)}

\newcommand\Xjm{\mathbf{X}_{jm}}
\newcommand\Zjm{\mathbf{Z}_{jm}}
\newcommand\Wjm{\mathbf{W}_{jm}}
\newcommand\Qjm{\mathbf{Q}_{jm}}

\newcommand\ajmc{\hat{a}_{jm}^\omega}
\newcommand\bjmc{\hat{b}_{jm}^\omega}

\newcommand\ajmx{\tilde{a}_{jm}^\omega}
\newcommand\bjmx{\tilde{b}_{jm}^\omega}

\newcommand\ajm{a_{jm}^\omega}
\newcommand\bjm{b_{jm}^\omega}
\newcommand\cjm{c_{jm}^\omega}
\newcommand\qjm{q_{jm}^\omega}

\newcommand\qpjm{g_{jm+}^{\omega}}
\newcommand\qmjm{g_{jm-}^\omega}
\newcommand\qzjm{g_{jm0}^\omega}
\newcommand\Qpjm{\mathbf{G}_{jm}^+}
\newcommand\Qmjm{\mathbf{G}_{jm}^-}
\newcommand\Qzjm{\mathbf{G}_{jm}^0}

\newcommand\mbar{\overline{m}}
\newcommand\lbar{\bar{l}}

\newcommand\intdOmegap{\int d \phat\text{ }}

\newcommand\intdr{\int d^3\rr\text{ }}

\newcommand\intdrprime{\int d^3\rr'\text{ }}
\newcommand\rhatprime{\mathbf{\hat{\rr}'}}
\newcommand\Aomegar{\mathbf{A}_\omega(\rr)}
\newcommand\Jomegarprime{\mathbf{J}_\omega(\rr')}
\newcommand\Jomegarprimeall{\mathbf{J}_\omega(\rr')}

\newcommand\dpp{\frac{d^3\pp}{\sqrt{(2\pi)^3}}}

\newcommand\Jone{{J}_1^\omega}
\newcommand\Jzero{{J}_0^\omega}
\newcommand\Jmone{{J}_{-1}^\omega}

\newcommand\factorfourier{\frac{1}{\sqrt{(2\pi)^3}}}

\newcommand\aoo{a_{11}^\omega}
\newcommand\boo{b_{11}^\omega}
\newcommand\coo{c_{11}^\omega}

\newcommand\aom{a_{1-1}^\omega}
\newcommand\bom{b_{1-1}^\omega}
\newcommand\com{c_{1-1}^\omega}

\newcommand\aoz{a_{10}^\omega}
\newcommand\boz{b_{10}^\omega}
\newcommand\coz{c_{10}^\omega}
 \begin{document}
\title{Exact dipolar moments of a localized electric current distribution}
\author{Ivan \surname{Fernandez-Corbaton}}
\email{ivan.fernandez-corbaton@kit.edu}
\affiliation{Institute of Nanotechnology, Karlsruhe Institute of Technology, 76021 Karlsruhe, Germany}
\author{Stefan Nanz}
\affiliation{Institut f\"ur Theoretische Festk\"orperphysik, Karlsruhe Institute of Technology, 76131 Karlsruhe, Germany}
\author{Rasoul Alaee}
\affiliation{Institut f\"ur Theoretische Festk\"orperphysik, Karlsruhe Institute of Technology, 76131 Karlsruhe, Germany}

\author{Carsten Rockstuhl}
\affiliation{Institut f\"ur Theoretische Festk\"orperphysik, Karlsruhe Institute of Technology, 76131 Karlsruhe, Germany}
\affiliation{Institute of Nanotechnology, Karlsruhe Institute of Technology, 76021 Karlsruhe, Germany}

\begin{abstract}
The multipolar decomposition of current distributions is used in many branches of physics. Here, we obtain new exact expressions for the dipolar moments of a localized electric current distribution. The typical integrals for the dipole moments of electromagnetically small sources are recovered as the lowest order terms of the new expressions in a series expansion with respect to the size of the source. All the higher order terms can be easily obtained. We also provide exact and approximated expressions for dipoles that radiate a definite polarization handedness (helicity). Formally, the new exact expressions are only marginally more complex than their lowest order approximations. 
 \end{abstract}
\maketitle
The multipolar decomposition of a spatially confined electromagnetic source distribution is a basic tool in both classical and quantum electrodynamics \cite{Landau1975,Jackson1998,Cohen1997,Craig1984,Walecka2004}. On the one hand, the multipolar coefficients determine the coupling of the source to external electromagnetic fields. This is used in the study of molecular, atomic, and nuclear electromagnetic interactions. On the other hand, there is a one-to-one correspondence between the multipolar components of the source and the multipolar fields radiated by it. This is exploited in the understanding and design of radiating systems. For example, in nanophotonics, the multipole moments of induced current distributions are used to study optical nano-antennas and meta-atoms \cite{Muhlig2011,Rockstuhl2011,Grahn2012,Arango2013}. The multipolar decomposition can be done in different ways, e.g. \cite[Chap. 9]{Jackson1998} and \cite[App. B, \S4]{Blatt1952}, resulting in integral expressions for the multipolar coefficients. The exact expressions are considerably simplified in the limit of electromagnetically small sources, but artificial scatterers at optical frequencies are typically large enough to compromise the accuracy of the approximation. 
\section{Outline}
In this article, we obtain new exact expressions for the source dipolar moments [Eqs. (\ref{eq:bexact})-(\ref{eq:cexact})]. In particular, they are valid for any source size.  We start our derivation in momentum space exploiting the fact that the fields radiated by the source at a given frequency $\omega$ are determined solely by its momentum components in a spherical shell of radius $\omega/c$, where $c$ is the speed of light in the medium. We first obtain hybrid integrals in momentum and coordinate space for all multipolar orders. In the dipolar case, we bring them to a form that is only marginally more complex than the typical integrals that give the dipolar moments of electromagnetically small sources. The additional complexity is the appearance of spherical Bessel functions. We identify the spherical Bessel functions as the elements that perform the necessary selection of the appropriate momentum shell. When the spherical Bessel functions are expanded around zero, the typical approximations for the magnetic and electric moments of electromagnetically small sources are recovered as the lowest order terms in the expansion. The toroidal dipole is recovered as the second term in the electric case. All higher order corrections are easily obtained as successive terms of the expansions. We include integral expressions for the magnetic corrections of order $k^3$ and the electric/toroidal corrections of order $k^4$. We also provide exact and approximated expressions for dipoles that radiate a definite polarization handedness (helicity) [Eq. (\ref{eq:helexact}) and Eq. (\ref{eq:helapprox})]. 

\section{Problem setting}
We start by considering an electric current density distribution $\Jrt$ embedded in an infinite, isotropic, and homogeneous medium characterized by real valued permittivity $\epsilon$ and permeability $\mu$. We assume $\Jrt$ to be confined in space so that $\Jrt=0$ for $|\rr|>R$. We consider its energy-momentum Fourier representation 

\begin{equation}
	\label{eq:firstb}
	\begin{split}
		&\Jrt=\mathcal{R}\left[\int_{0^{+}}^{\infty}\frac{d\omega}{\sqrt{2\pi}}\exp\left(-i\omega t\right)\mathbf{J}_\omega(\rr)\right]\\
			   &=\mathcal{R}\left[\int_{0^{+}}^{\infty}\frac{d\omega}{\sqrt{2\pi}}\exp\left(-i\omega t\right)\int \frac{d^3\pp}{\sqrt{(2\pi)^3}}\ \Jomegagen\exp\left(i\pp\cdot\rr\right)\right],
	\end{split}
 \end{equation}
 and treat each $\omega$ term separately. The frequency $\omega$ and the three components of the momentum vector $\pp$ are real numbers. The lower limit of the integral in $d\omega$ excludes the static case $\omega=0$, which we do not treat in this paper. At each frequency $\omega$, the transverse electromagnetic fields outside the source are solely determined by the part of $\Jomegagen$ in the domain that satisfies $|\pp|=\omega/c$.  This result was obtained by Devaney and Wolf \cite{Devaney1974}. We provide an alternative proof in App. \ref{sec:onlyw}. 
 
We denote by $\Jomega$ the components of $\mathbf{J}_\omega(\pp)$ in the spherical shell of radius $|\pp|=\omega/c$. The symbol $\phat$ represents the angular part of the momentum vector $\pp$, i.e., the solid angle in the spherical shell. As usual, we define $k=\omega/c$.

We will expand $\Jomega$ in an orthonormal basis for functions defined in a spherical shell: The three families of multipolar functions in momentum space \cite[B$_I$.3]{Cohen1997}
\begin{equation}
	\label{eq:xzw}
	\begin{split}
		\Xjm(\phat)&=\frac{1}{\sqrt{j(j+1)}} \mathbf{L}Y_{jm}(\phat),\\
		\Zjm(\phat)&=i\phat\times\Xjm(\phat),\\ 
		\Wjm(\phat)&=\phat Y_{jm}(\phat).
	\end{split}
\end{equation}
The $Y_{jm}(\phat)$ are the spherical harmonics and the three components of the vector $\mathbf{L}$ are the angular momentum operators for scalar functions.

Each of the vector multipolar functions in the three families is an eigenstate of the total angular momentum squared $J^2$ and the angular momentum along one axis $\mathbf{\hat{q}}$, for which we choose $\mathbf{\hat{q}}=\zhat$. With $\Qjm(\phat)$ standing for any of the $\{\Xjm(\phat),\Zjm(\phat),\Wjm(\phat)\}$:
\begin{equation}
	J^2\Qjm(\phat)=j(j+1)\Qjm(\phat),\ J_z\Qjm(\phat)=m\Qjm(\phat), 
\end{equation}
where $j$ and $m$ are integers, and $m=-j\ldots j$.  For $\Xjm(\phat)$ and $\Zjm(\phat)$, $j$ takes integer values in $j>0$, while for $\Wjm(\phat)$, $j=0$ is also possible.

The functions in Eq. (\ref{eq:xzw}) are also eigenstates of the parity operator\footnote{Their eigenvalues can be deduced from the parity transformation properties of a vector field in momentum space, i.e. $\Pi \mathbf{F}(\pp) = -\mathbf{F}(-\pp)$, and those of the spherical harmonics, $\Pi Y_{jm}(\phat)=Y_{jm}(-\phat)=(-1)^jY_{jm}(\phat)$.}:

\begin{equation}
	\label{eq:parity}
	\begin{split}
		\Pi \Xjm(\phat) &= - \Xjm(-\phat)=(-1)^{j+1} \Xjm(\phat),\\
		\Pi \Zjm(\phat) &= - \Zjm(-\phat)=(-1)^{j} \Zjm(\phat),\\
		\Pi \Wjm(\phat) &= - \Wjm(-\phat)=(-1)^{j} \Wjm(\phat).
	\end{split}
\end{equation}

\begin{figure}[ht!]
\begin{center}
	\def\ASYprefix{}
\newbox\ASYbox
\newdimen\ASYdimen
\long\def\ASYbase#1#2{\leavevmode\setbox\ASYbox=\hbox{#1}\ASYdimen=\ht\ASYbox\setbox\ASYbox=\hbox{#2}\lower\ASYdimen\box\ASYbox}
\long\def\ASYaligned(#1,#2)(#3,#4)#5#6#7{\leavevmode\setbox\ASYbox=\hbox{#7}\setbox\ASYbox\hbox{\ASYdimen=\ht\ASYbox\advance\ASYdimen by\dp\ASYbox\kern#3\wd\ASYbox\raise#4\ASYdimen\box\ASYbox}\put(#1,#2){#5\wd\ASYbox 0pt\dp\ASYbox 0pt\ht\ASYbox 0pt\box\ASYbox#6}}\long\def\ASYalignT(#1,#2)(#3,#4)#5#6{\ASYaligned(#1,#2)(#3,#4){
}{
}{#6}}
\long\def\ASYalign(#1,#2)(#3,#4)#5{\ASYaligned(#1,#2)(#3,#4){}{}{#5}}
\def\ASYraw#1{
currentpoint currentpoint translate matrix currentmatrix
100 12 div -100 12 div scale
#1
setmatrix neg exch neg exch translate}
 	\includegraphics[width=0.65\linewidth]{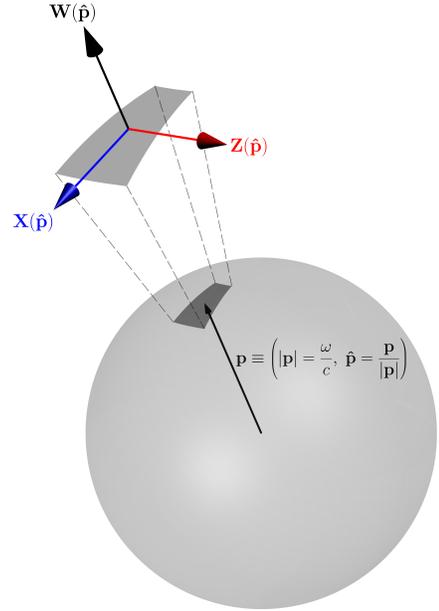}
	\caption{\label{fig:xzw} The electromagnetic field radiated by a confined monochromatic current density $\Jomegar$ with Fourier transform $\mathbf{J}_\omega(\pp)$ only depends on the components of $\mathbf{J}_\omega(\pp)$ in a spherical shell of radius $|\pp|=\omega/c$. The relevant part of $\mathbf{J}_\omega(\pp)$ can hence be expressed as a linear combination of the  momentum space vector multipolar functions $\{\Xjm(\phat),\Zjm(\phat),\Wjm(\phat)\}$, which form an orthonormal basis for functions defined on the shell. The polarization vectors of $\Xjm(\phat)$ and $\Zjm(\phat)$ are tangential to the surface of the shell, i.e., orthogonal (transverse) to the momentum vector $\pp$. The polarization vector of $\Wjm(\phat)$ is normal to the surface of the shell, i.e., parallel (longitudinal) to $\pp$.} 
\end{center}
\end{figure}

The polarization of $\Xjm(\phat)$ and $\Zjm(\phat)$ is transverse (orthogonal) to $\phat$, and the polarization of $\Wjm(\phat)$ is longitudinal (parallel) to $\phat$, as depicted in Fig. \ref{fig:xzw}. In coordinate ($\rr$) space, this distinction corresponds to the distinction between divergence free (transverse) and curl free (longitudinal) fields.

With the scalar product
\begin{equation}
\langle A | B \rangle = \intdOmegap \mathbf{A}^\dagger(\phat)\mathbf{B}(\phat),
\end{equation}
where $^{\dagger}$ denotes hermitian transpose, and $\phat$ runs over the entire spherical shell, the three families together form an orthonormal basis for functions defined on any spherical shell in momentum space. 

We expand $\Jomega$ in this basis:
\begin{equation}
	\label{eq:expand}
	\Jomega=\sum_{jm} \ajm\Zjm(\phat)+\bjm\Xjm(\phat)+\cjm\Wjm(\phat),
\end{equation}
where, with $\qjm$ standing for any of the $\{\ajm,\bjm,\cjm\}$,
\begin{equation}
	\label{eq:qjm}
	\qjm=\langle Q_{jm}|\mathring{J}_\omega\rangle=\int d\phat \Qjm^\dagger(\phat) \Jomega.
\end{equation}

The $\{\ajm,\bjm,\cjm\}$ coefficients contain all the information about $\Jomega$ so they must also contain all the information about the fields produced by it. As shown in \cite{Devaney1974}, the $\{\ajm,\bjm\}$ determine the transverse electromagnetic field radiated by the sources at frequency $\omega$ outside a spherical volume enclosing them: They are the coefficients of the expansion of the transverse fields in outgoing electric and magnetic multipoles, respectively \cite[Eq. 9.122]{Jackson1998}. Therefore, the transverse components of $\Jomega$ determine the transverse components of the electromagnetic field at frequency $\omega$ outside the source region. The longitudinal electric field with $|\pp|=\omega/c$ is zero outside the source region. While the longitudinal degrees of freedom of $\Jomega$, i.e. the $\cjm$, are not necessarily equal to zero, the field that they generate outside the source region is canceled by the field generated by the charge density. This can be seen in \cite[\S 13.3 p1875-1877]{Morse1953}, and in \cite[App. C]{FerCor2015c} where the cancellation is shown to be a consequence of the continuity equation. We will keep the $\cjm$ in the discussion both for completeness and because they play an important role in understanding the split of the $\ajm$ into electrical and toroidal parts \cite{Dubovik1974,Radescu2002,Kaelberer2010}, which we discuss in \cite{FerCor2015c}. 

The $\{\ajm,\bjm\}$ coefficients are a valuable source of information in many branches of physics. In molecular, atomic and nuclear physics, the $\{\ajm,\bjm\}$ coefficients are used to describe the interaction of systems of charges with external electromagnetic fields, e.g. \cite[Chap. 10]{Craig1984}, \cite[IV.C.2c)]{Cohen1997} and \cite[Chap. 7]{Walecka2004}. In classical electrodynamics they are used to describe radiation by source distributions, e.g. \cite[Chap. 9]{Landau1975} and \cite[Chap. 9]{Jackson1998}. In nanophotonics, they are used to study and design the response of individual artificial nanostructures. 

Given $\Jomegar$, there exist exact expressions for the $\{\ajm,\bjm\}$ as coordinate space integrals, e.g. \cite[Eq. (7.20)]{Walecka2004}

\begin{equation}
	\label{eq:wal}
	\begin{split}
		\ajmx&=\frac{1}{k}\int d^3\rr \ \left(\nabla\times j_j(kr)\Xjm(\rhat)\right)^\dagger \Jomegar,\\
		\bjmx&=\int d^3\rr \ j_j(kr)\mathbf{X}_{jm}^\dagger(\rhat) \Jomegar,
	\end{split}
\end{equation}
or, \cite[Eq. (9.165) without the magnetization current therein]{Jackson1998},
\begin{equation}
	\label{eq:jac}
	\begin{split}
		\ajmc&=\frac{ik}{\sqrt{j(j+1)}}\int d^3\rr \ j_j(kr)Y^*_{jm}(\rhat)\mathbf{L}\cdot(\nabla\times\Jomegar),\\
		\bjmc&=\frac{-k^2}{\sqrt{j(j+1)}}\int d^3\rr \ j_j(kr)Y^*_{jm}(\rhat)\mathbf{L}\cdot\Jomegar,
	\end{split}
\end{equation}
where the tildes and carets in the left hand sides indicate different normalizations, $r=|\rr|$, $\rhat=\rr/|\rr|$ is the angular part of $\rr$, and $k\equiv \omega/c$ throughout the article. 

The expressions in Eq. (\ref{eq:wal}) and Eq. (\ref{eq:jac}) are valid for any source radius $R$. For electromagnetically small sources where $kR\ll1$, they can be reduced to the simpler well known expressions that are obtained in \cite[Chap. 9]{Jackson1998} and \cite[Chap. 9]{Landau1975} by starting with the equation for the vector potential as a function of $\Jomegar$ in the Lorentz gauge [Eq. (\ref{eq:first})], and expanding $\frac{\exp\left({ik|\rr-\rr'|}\right)}{|\rr-\rr'|}$ in powers of $k|\rr-\rr'|$. For example, when $kR\ll1$ the source dependent terms of the electric and magnetic dipole moments are 

\begin{equation}
\begin{bmatrix}\aoo\\\aoz\\\aom\end{bmatrix}\rightarrow\intdr \Jomegar,\ \begin{bmatrix}\boo\\\boz\\\bom\end{bmatrix}\rightarrow\intdr \rr\times\Jomegar,\\
\end{equation}
where we have chosen the spherical vector basis. We will work in this basis throughout the article. Appendix \ref{sec:supporting} contains auxiliary expressions.

\section{Exact dipolar moments\label{sec:edm}}
We will now obtain exact expressions for the dipolar vectors $[\aoo,\aoz,\aom]^T$, $[\boo,\boz,\bom]^T$ and $[\coo,\coz,\com]^T$ as coordinate space integrals of functions of $\Jomegar$. While these expressions are, as Eq. (\ref{eq:wal}) and Eq. (\ref{eq:jac}), valid for any source size, they are only marginally more complex than their $kR\ll1$ limits: Namely, they contain spherical Bessel functions. As far as we know, these expressions have not been reported before. 

We start from Eq. (\ref{eq:qjm}), where we substitute
\begin{equation}
	\label{eq:jwp}
		\Jomega=\factorfourier\intdr \Jomegar \exp\left(-i\frac{\omega}{c}\phat\cdot\rr\right).
\end{equation}
to get
\begin{equation}
	\label{eq:jwp2}
	\qjm=\factorfourier\intdOmegap \Qjm^\dagger(\phat) \intdr \Jomegar \exp\left(-i\frac{\omega}{c}\phat\cdot\rr\right).
\end{equation}
The condition $|\pp|=\omega/c$ is enforced in the argument of the exponential. We now substitute the exponential for its expansion in spherical harmonics
\begin{equation}
	\label{eq:exp}
	\exp\left(-i\frac{\omega}{c}\phat\cdot\rr\right)=(4\pi)\sum_{\lbar,\mbar}(-i)^{\lbar} Y^*_{\lbar\mbar}(\rhat)Y_{\lbar\mbar}(\phat)j_{\lbar}(k|\rr|),
\end{equation}
where $j_{\lbar}(\cdot)$ is the $\lbar$-th order spherical Bessel function of the first kind. The result is:
\begin{equation}
	\label{eq:exact}
	\begin{split}
	&\frac{\sqrt{(2\pi)^3}}{4\pi}\qjm=\\
	 &\sum_{\lbar\mbar}(-i)^{\lbar}\shadedbox{\intdOmegap \Qjm^\dagger(\phat) Y_{\lbar\mbar}(\phat)} \intdr \Jomegar Y^*_{\lbar\mbar}(\rhat)j_{\lbar}(kr).
\end{split}
\end{equation}

Equation (\ref{eq:exact}) is an exact expression for the $\{\ajm,\bjm,\cjm\}$ coefficients in terms of integrals in both momentum (shaded area) and coordinate space. As shown in App. \ref{sec:omte}, only terms with $\lbar=j$ contribute to the $\bjm$, while the $\ajm$ and $\cjm$ get contributions from both $\lbar=j-1$ and $\lbar=j+1$. Additionally, it is possible to further simplify Eq. (\ref{eq:exact}) in the dipolar ($j=1$) case without making any approximation. We now present the derivations for the magnetic dipole $b_{1m}^\omega$. Appendix \ref{sec:appB} contains the derivations for $a_{1m}^\omega$ and $c_{1m}^\omega$. It also contains the $c_{00}$ case.

For the magnetic dipole, we particularize Eq. (\ref{eq:exact}) for $\Qjm(\phat)\rightarrow\Xjm(\phat)$ and $j=1$, which implies $\lbar=1$:

\begin{equation}
	\label{eq:exactb}
	\begin{split}
		&i\frac{\sqrt{(2\pi)^3}}{4\pi}b_{1m}^\omega=\\
		&\sum_{\mbar=-1}^{\mbar=1}\shadedbox{\intdOmegap \mathbf{X}_{1m}^\dagger(\phat) Y_{1\mbar}(\phat)}{\intdr {\Jomegar} Y^*_{1\mbar}(\rhat)j_{1}(kr)}.
	\end{split}
\end{equation}

Explicit expressions of $\mathbf{X}_{1m}(\phat)$ can be obtained using Eq. (\ref{eq:xjm}) and then used to write the momentum integrals in the shaded area of Eq. (\ref{eq:exactb}) as 
 \begin{equation}
	 \begin{split}
	 &m=1\rightarrow\ \intdOmegap\begin{bmatrix}-\frac{Y_{10}(\phat)}{\sqrt{2}}\\\frac{Y_{11}(\phat)}{\sqrt{2}}\\0\end{bmatrix}^\dagger Y_{1\mbar}(\phat),\\
	 &m=0\rightarrow \intdOmegap\begin{bmatrix}-\frac{Y_{1-1}(\phat)}{\sqrt{2}}\\0\\\frac{Y_{11}(\phat)}{\sqrt{2}}\end{bmatrix}^\dagger Y_{1\mbar}(\phat),\\
	 &m=-1\rightarrow\ \intdOmegap\begin{bmatrix}0\\-\frac{Y_{1-1}(\phat)}{\sqrt{2}}\\\frac{Y_{10}(\phat)}{\sqrt{2}}\end{bmatrix}^\dagger Y_{1\mbar}(\phat),
	 \end{split}
 \end{equation}
 which can be easily solved for each $\mbar\in\{-1,0,1\}$ using the orthonormality properties of the spherical harmonics: $\intdOmegap Y_{lm}(\phat)Y_{\lbar\mbar}^*(\phat)=\delta_{\mbar m}\delta_{\lbar l}$. They result in three vectors for each $m$ case, which we list here as row vectors. From top to bottom, the three row vectors correspond to $\mbar=1,0,-1$:
\begin{equation}
	\begin{split}
	m=1&\rightarrow\ \frac{1}{\sqrt{2}}\begin{array}{ccccc}(&0&1&0&)\\(&-1&0&0&),\\(&0&0&0&)\end{array}\\
	m=0&\rightarrow\ \frac{1}{\sqrt{2}}\begin{array}{ccccc}(&0&0&1&)\\(&0&0&0&),\\(&-1&0&0&)\end{array}\\
	m=-1&\rightarrow\ \frac{1}{\sqrt{2}}\begin{array}{ccccc}(&0&0&0&)\\(&0&0&1&).\\(&0&-1&0&)\end{array}
	\end{split}
\end{equation}
Having solved the momentum space integrals in the shaded area of Eq. (\ref{eq:exactb}), the summation in $\mbar$ can now be done. With $\Jomegar=[\Jone ,\Jzero ,\Jmone ]^T$, and, as in Eq. (\ref{eq:phat}),
\begin{equation}
\rhat=\frac{\rr}{|\rr|}=2\sqrt{\frac{\pi}{3}}\begin{bmatrix}Y_{11}^*(\rhat)\\Y_{10}^*(\rhat)\\Y_{1-1}^*(\rhat)\end{bmatrix},
	\end{equation}
the result of the sum reads
\begin{equation}
	\label{eq:m}
	\begin{split}
		\boo&=\frac{\sqrt{3}}{2\pi i}\intdr \left(\Jzero \hat{r}_{1}-\Jone \hat{r}_{0}\right)j_1(kr),\\
		\boz&=\frac{\sqrt{3}}{2\pi i}\intdr \left(\Jmone \hat{r}_{1}-\Jone \hat{r}_{-1}\right)j_1(kr),\\
		\bom&=\frac{\sqrt{3}}{2\pi i}\intdr \left(\Jmone \hat{r}_{0}-\Jzero \hat{r}_{-1}\right)j_1(kr).\\
	\end{split}
\end{equation}
Considering the expression for the cross product in spherical coordinates [Eq. (\ref{eq:cross})], we can finally write Eq. (\ref{eq:m}) as:
\begin{equation}
	\label{eq:bexact}
\begin{bmatrix}\boo\\\boz\\\bom\end{bmatrix}=-\frac{\sqrt{3}}{2\pi}\intdr  \rhat\times\Jomegar j_1(kr).
\end{equation}

The expressions for $[\aoo,\aoz,\aom]^T$ and $[\coo,\coz,\com]^T$ can be obtained by similar, although more involved, procedures. We provide the derivations in App. \ref{sec:appB}. The results read:
\begin{equation}
	\label{eq:aexact}
		\begin{split}
		\begin{bmatrix}\aoo\\\aoz\\\aom\end{bmatrix}&=\underbrace{-\frac{1}{\pi\sqrt{3}}\intdr  \Jomegar j_0(kr)}_{\lbar=0}\\
	 &\underbrace{-\frac{1}{2\pi\sqrt{3}}\intdr\left\{3\left[\rhat^{\dagger}\Jomegar\right]\rhat-\Jomegar\right\}j_2(kr)}_{\lbar=2},
		\end{split}
\end{equation}
and
\begin{equation}
	\label{eq:cexact}
		\begin{split}
		\begin{bmatrix}\coo\\\coz\\\com\end{bmatrix}&=\underbrace{\frac{1}{\pi\sqrt{6}}\intdr \Jomegar j_0(kr)}_{\lbar=0}\\
	   &\underbrace{-\frac{1}{\pi\sqrt{6}}\intdr\left\{3\left[\rhat^{\dagger}\Jomegar\right]\rhat-\Jomegar\right\}j_2(kr)}_{\lbar=2},
		\end{split}
\end{equation}
where the contributions coming from $\lbar=j-1=0$ and $\lbar=j+1=2$ are indicated. The dot product $\rhat^{\dagger}\Jomegar$ is simply equal to $\rhat^{T}\Jomegar$ in Cartesian coordinates\footnote{This can be seen using the complex unitary matrix that transforms spherical into Cartesian components, written on the right of Eq. (\ref{eq:cartsph}). Calling such matrix $C$, we start with the dot product in spherical coordinates and transform it to Cartesian coordinates using that $(C^{\dagger} C)$ is the identity: $\rhat^\dagger\Jomegar=\rhat^\dagger(C^{\dagger} C)\Jomegar=(C\rhat)^\dagger(C\Jomegar)=\rhat_{cart}^\dagger\Jomegar_{cart}=\rhat_{cart}^T\Jomegar_{cart}$, where the last equality follows because the Cartesian coordinates $(x,y,z)$ are real.}. 

Equation (\ref{eq:bexact}), Eq. (\ref{eq:aexact}), and Eq. (\ref{eq:cexact}) are exact. In particular they apply to a source distribution of any size. They are also simpler than the corresponding exact expressions obtained from Eq. (\ref{eq:wal}) or Eq. (\ref{eq:jac}). We note that Eqs. (\ref{eq:bexact}) to Eq. (\ref{eq:cexact}) should also be reachable from the coordinate space integrals of Eq. (\ref{eq:wal}) or Eq. (\ref{eq:jac}). Our route through momentum space explicitly exploits that the contributions to the $\qjm$ only come from the Fourier components of the source in the domain $|\pp|=\omega/c$. This restriction is imposed in the exponential of Eq. (\ref{eq:jwp}) and determines the argument of the spherical Bessel functions $j_{\lbar}(kr)$ in Eq. (\ref{eq:exact}), which then appear in Eqs. (\ref{eq:bexact}), Eq. (\ref{eq:aexact}), and Eq. (\ref{eq:cexact}). We can deduce that the spherical Bessel functions must be responsible for rejecting the $|\pp|\neq\omega/c$ components present in $\Jomegar$. We now provide a more formal proof of their role.

In the expression of $\qjm$ in Eq. (\ref{eq:exact}), the dependence on the current density is contained in the integrals
\begin{equation}
	\intdr \Jomegar Y^*_{lm}(\rhat)j_{l}(kr).
\end{equation}
We then write $\Jomegar$ as an inverse Fourier transform and expand its exponential $\exp\left(i\pp\cdot\rr\right)$ as in Eq. (\ref{eq:exp}), except that now $|\pp|$ {\em is not restricted to $\omega/c$}. After rearranging the integrals we get:
{\small
	\begin{equation}
\label{eq2:aux}
	\sum_{\lbar,\mbar}	\frac{4\pi i^{\lbar}}{\sqrt{(2\pi)^3}}\int d^3\pp \mathbf{J}_\omega(\pp)Y^*_{\lbar\mbar}(\phat)\shadedbox{\int d^3\rr Y_{\lbar\mbar}(\rhat)Y^*_{lm}(\rhat)j_{\lbar}(|\pp|r)j_{l}(kr)}.
	\end{equation}
}
The shaded $d^3\rr$ integral can be solved by splitting it into its radial and angular parts $\left(\int d^3\rr=\int_0^{\infty}dr\ r^2\int d\rhat\right)$. First, the angular part is solved through the orthonormality of the spherical harmonics, which forces $(\lbar,\mbar)=(l,m)$. The remaining radial integral has a formal solution as a radial Dirac delta distribution \cite[Eq. (4.1)]{Mehrem1991}
\begin{equation}
	\label{eq:delta2}
	\int drr^2  j_l(|\pp|r)j_l(kr) =\frac{\pi}{2k^2}\delta(|\pp|-k),
\end{equation}
	which enforces the $|\pp|=k=\omega/c$ restriction in Eq. (\ref{eq2:aux}), namely:
\begin{equation}
	\begin{split}
		\frac{4\pi i^{l}}{\sqrt{(2\pi)^3}}&\int \ d^3\pp \mathbf{J}_\omega(\pp)Y^*_{lm}(\phat)\frac{\pi}{2k^2}\delta(|\pp|-k)=\\
		\frac{4\pi i^{l}}{\sqrt{(2\pi)^3}}&\int \ d\phat Y^*_{lm}(\phat)\int_0^{\infty}dp\ p^2 \mathbf{J}_\omega(\pp)\frac{\pi}{2k^2}\delta(|\pp|-k)=\\
		\frac{1}{k^2}\sqrt{\frac{\pi}{2}}i^{l}&\int \ d\phat \Jomega Y^*_{lm}(\phat).
	\end{split}
\end{equation}

The $j_l(kr)$ functions from Eq. (\ref{eq:exact}) find their way into Eq. (\ref{eq:delta2}), and become one of the pieces needed to obtain the Dirac delta $\delta(|\pp|-k)$ which filters out the $|\pp|\neq \omega/c$ components of $\Jomegar$.

\section{Electromagnetically small source approximation with increasing accuracy}
We now make the small argument approximation to the spherical Bessel functions in Eqs. (\ref{eq:bexact})-(\ref{eq:cexact}) and keep terms up to second order: $j_0(kr)\approx 1-(kr)^2/6$, $j_1(kr)\approx kr/3$ and $j_2(kr)\approx (kr)^2/15$. After grouping terms with the same power of $k$ we obtain:
\begin{eqnarray}
\label{eq:bapprox}
&\begin{bmatrix}\boo\\\boz\\\bom\end{bmatrix}\approx -\frac{1}{2\pi\sqrt{3}}k \intdr\ \rr\times\Jomegar,\\
\label{eq:eapprox}
&\begin{bmatrix} \aoo\\\aoz\\\aom\end{bmatrix}\approx -\underbrace{\frac{1}{\pi\sqrt{3}}\intdr \Jomegar}_{\lbar=0}\\
\label{eq:tapprox}
&\underbrace{-\frac{1}{\pi\sqrt{3}}k^2\intdr\frac{1}{10}\left\{\left[\rr^{\dagger}\Jomegar\right]\rr-2r^2\Jomegar\right\}}_{\lbar=0,\lbar=2},\\
\label{eq:ceapprox}
  &\begin{bmatrix}\coo\\\coz\\\com\end{bmatrix}\approx\underbrace{\frac{1}{\pi\sqrt{6}}\intdr \Jomegar}_{\lbar=0}\\
\label{eq:ctapprox}
&\underbrace{-\frac{1}{\pi}\sqrt{\frac{2}{3}}k^2\intdr\frac{1}{10}\left\{2\left[\rr^{\dagger}\Jomegar\right]\rr+r^2\Jomegar\right\}}_{\lbar=0,\lbar=2}.
\end{eqnarray}

Equation (\ref{eq:bapprox}), Eq. (\ref{eq:eapprox}) and Eq. (\ref{eq:tapprox}) are, respectively, the well known approximated magnetic, electric, and toroidal dipole moments of electromagnetically small current distributions. We note that the electric dipole contains contributions only from $\lbar=0$ while the toroidal dipole has contributions from $\lbar=0$ and $\lbar=2$. 

The small argument approximation causes two kinds of inaccuracies. On the one hand, entire integral terms are neglected. For example, the toroidal term in Eq. (\ref{eq:tapprox}) disappears in a lowest order approximation. On the other hand, some components with $|\pp|\neq\omega/c$ will leak into the dipole moments. This happens because the approximated expressions of the spherical Bessel functions do not correspond to momentum space Dirac deltas $\delta(|\pp|-k)$.

Approximations with increasing accuracy are obtained in a straightforward way from the exact Eqs. (\ref{eq:bexact}) to Eq. (\ref{eq:cexact}). It is a matter of taking more terms in the expansions of the spherical Bessel functions. For example, the $(kr)^3$ correction to Eq. (\ref{eq:bapprox}) reads 

\begin{equation}
	\label{eq:k3corr}
	\frac{\sqrt{3}k^3}{60\pi}\intdr \left[\rr\times\Jomegar\right] r^2,
\end{equation}
the $(kr)^4$ correction to the total $a_{1m}^\omega$ in Eqs. (\ref{eq:eapprox})-(\ref{eq:tapprox}) reads 
\begin{equation}
	\label{eq:ak4corr}
	\frac{k^4}{140\pi\sqrt{3}}\intdr \left\{\left[\rr^{\dagger}\Jomegar\right]\rr-\frac{3}{2}\Jomegar r^2\right\}r^2,
\end{equation}
and the $(kr)^4$ correction to the total $c_{1m}^\omega$ in Eqs. (\ref{eq:ceapprox})-(\ref{eq:ctapprox}) reads 
\begin{equation}
	\frac{k^4}{70\pi\sqrt{6}}\intdr \left\{\left[\rr^{\dagger}\Jomegar\right]\rr+\frac{1}{4}\Jomegar r^2\right\}r^2.
\end{equation}

The above corrections to $a_{1m}^\omega$ and $b_{1m}^\omega$ coincide up to normalization factors with the {\em mean square radii} in \cite[App. C]{Radescu2002}, where they are derived in a different way.

We now use our results to compute the magnetic dipole moment of a current distribution with a previously known analytical solution, verify that the result coincides, and compare it with two approximated solutions for electromagnetically small sources obtained from taking the first and the two first terms in the expansion of the spherical Bessel functions.

\section{Example}
\begin{figure}[ht]
\begin{center}
\begingroup
  \makeatletter
  \providecommand\color[2][]{    \GenericError{(gnuplot) \space\space\space\@spaces}{      Package color not loaded in conjunction with
      terminal option `colourtext'    }{See the gnuplot documentation for explanation.    }{Either use 'blacktext' in gnuplot or load the package
      color.sty in LaTeX.}    \renewcommand\color[2][]{}  }  \providecommand\includegraphics[2][]{    \GenericError{(gnuplot) \space\space\space\@spaces}{      Package graphicx or graphics not loaded    }{See the gnuplot documentation for explanation.    }{The gnuplot epslatex terminal needs graphicx.sty or graphics.sty.}    \renewcommand\includegraphics[2][]{}  }  \providecommand\rotatebox[2]{#2}  \@ifundefined{ifGPcolor}{    \newif\ifGPcolor
    \GPcolortrue
  }{}  \@ifundefined{ifGPblacktext}{    \newif\ifGPblacktext
    \GPblacktexttrue
  }{}    \let\gplgaddtomacro\g@addto@macro
    \gdef\gplbacktext{}  \gdef\gplfronttext{}  \makeatother
  \ifGPblacktext
        \def\colorrgb#1{}    \def\colorgray#1{}  \else
        \ifGPcolor
      \def\colorrgb#1{\color[rgb]{#1}}      \def\colorgray#1{\color[gray]{#1}}      \expandafter\def\csname LTw\endcsname{\color{white}}      \expandafter\def\csname LTb\endcsname{\color{black}}      \expandafter\def\csname LTa\endcsname{\color{black}}      \expandafter\def\csname LT0\endcsname{\color[rgb]{1,0,0}}      \expandafter\def\csname LT1\endcsname{\color[rgb]{0,1,0}}      \expandafter\def\csname LT2\endcsname{\color[rgb]{0,0,1}}      \expandafter\def\csname LT3\endcsname{\color[rgb]{1,0,1}}      \expandafter\def\csname LT4\endcsname{\color[rgb]{0,1,1}}      \expandafter\def\csname LT5\endcsname{\color[rgb]{1,1,0}}      \expandafter\def\csname LT6\endcsname{\color[rgb]{0,0,0}}      \expandafter\def\csname LT7\endcsname{\color[rgb]{1,0.3,0}}      \expandafter\def\csname LT8\endcsname{\color[rgb]{0.5,0.5,0.5}}    \else
            \def\colorrgb#1{\color{black}}      \def\colorgray#1{\color[gray]{#1}}      \expandafter\def\csname LTw\endcsname{\color{white}}      \expandafter\def\csname LTb\endcsname{\color{black}}      \expandafter\def\csname LTa\endcsname{\color{black}}      \expandafter\def\csname LT0\endcsname{\color{black}}      \expandafter\def\csname LT1\endcsname{\color{black}}      \expandafter\def\csname LT2\endcsname{\color{black}}      \expandafter\def\csname LT3\endcsname{\color{black}}      \expandafter\def\csname LT4\endcsname{\color{black}}      \expandafter\def\csname LT5\endcsname{\color{black}}      \expandafter\def\csname LT6\endcsname{\color{black}}      \expandafter\def\csname LT7\endcsname{\color{black}}      \expandafter\def\csname LT8\endcsname{\color{black}}    \fi
  \fi
  \setlength{\unitlength}{0.0500bp}  \begin{picture}(5040.00,3968.00)    \gplgaddtomacro\gplbacktext{      \csname LTb\endcsname      \put(594,704){\makebox(0,0)[r]{\strut{} 0}}      \put(594,1132){\makebox(0,0)[r]{\strut{} 10}}      \put(594,1561){\makebox(0,0)[r]{\strut{} 20}}      \put(594,1989){\makebox(0,0)[r]{\strut{} 30}}      \put(594,2418){\makebox(0,0)[r]{\strut{} 40}}      \put(594,2846){\makebox(0,0)[r]{\strut{} 50}}      \put(594,3275){\makebox(0,0)[r]{\strut{} 60}}      \put(594,3703){\makebox(0,0)[r]{\strut{} 70}}      \put(726,484){\makebox(0,0){\strut{} 0}}      \put(1286,484){\makebox(0,0){\strut{} 0.1}}      \put(1845,484){\makebox(0,0){\strut{} 0.2}}      \put(2405,484){\makebox(0,0){\strut{} 0.3}}      \put(2964,484){\makebox(0,0){\strut{} 0.4}}      \put(3524,484){\makebox(0,0){\strut{} 0.5}}      \put(4083,484){\makebox(0,0){\strut{} 0.6}}      \put(4643,484){\makebox(0,0){\strut{} 0.7}}      \put(220,2203){\rotatebox{-270}{\makebox(0,0){\strut{}Percentage of relative error}}}      \put(1628,154){\makebox(0,0){\large $\frac{2a}{\lambda_0}$}\hspace{0.4cm}  (diameter over wavelength)}    }    \gplgaddtomacro\gplfronttext{      \csname LTb\endcsname      \put(3005,3413){\makebox(0,0)[r]{\strut{}$100\times\left|1-\frac{ka}{3j_1(ka)}\right|$}}      \csname LTb\endcsname      \put(3005,2961){\makebox(0,0)[r]{\strut{}$100\times\left|1-\frac{ka\left[1-(ka)^2/10\right]}{3j_1(ka)}\right|$}}      \csname LTb\endcsname      \put(2293,2375){\makebox(0,0)[l]{\strut{}$a$}}      \put(1845,1861){\makebox(0,0)[l]{\strut{}$\Jomegar$}}      \put(2125,1475){\makebox(0,0){\strut{}Current loop}}    }    \gplgaddtomacro\gplbacktext{      \put(594,704){\makebox(0,0)[r]{\strut{} 0}}      \put(594,1132){\makebox(0,0)[r]{\strut{} 10}}      \put(594,1561){\makebox(0,0)[r]{\strut{} 20}}      \put(594,1989){\makebox(0,0)[r]{\strut{} 30}}      \put(594,2418){\makebox(0,0)[r]{\strut{} 40}}      \put(594,2846){\makebox(0,0)[r]{\strut{} 50}}      \put(594,3275){\makebox(0,0)[r]{\strut{} 60}}      \put(594,3703){\makebox(0,0)[r]{\strut{} 70}}      \put(726,484){\makebox(0,0){\strut{} 0}}      \put(1286,484){\makebox(0,0){\strut{} 0.1}}      \put(1845,484){\makebox(0,0){\strut{} 0.2}}      \put(2405,484){\makebox(0,0){\strut{} 0.3}}      \put(2964,484){\makebox(0,0){\strut{} 0.4}}      \put(3524,484){\makebox(0,0){\strut{} 0.5}}      \put(4083,484){\makebox(0,0){\strut{} 0.6}}      \put(4643,484){\makebox(0,0){\strut{} 0.7}}      \put(220,2203){\rotatebox{-270}{\makebox(0,0){\strut{}Percentage of relative error}}}      \put(1628,154){\makebox(0,0){\large $\frac{2a}{\lambda_0}$}\hspace{0.4cm}  (diameter over wavelength)}    }    \gplgaddtomacro\gplfronttext{      \csname LTb\endcsname      \put(2293,2375){\makebox(0,0)[l]{\strut{}$a$}}      \put(1845,1861){\makebox(0,0)[l]{\strut{}$\Jomegar$}}      \put(2125,1475){\makebox(0,0){\strut{}Current loop}}    }    \gplgaddtomacro\gplbacktext{      \put(594,704){\makebox(0,0)[r]{\strut{} 0}}      \put(594,1132){\makebox(0,0)[r]{\strut{} 10}}      \put(594,1561){\makebox(0,0)[r]{\strut{} 20}}      \put(594,1989){\makebox(0,0)[r]{\strut{} 30}}      \put(594,2418){\makebox(0,0)[r]{\strut{} 40}}      \put(594,2846){\makebox(0,0)[r]{\strut{} 50}}      \put(594,3275){\makebox(0,0)[r]{\strut{} 60}}      \put(594,3703){\makebox(0,0)[r]{\strut{} 70}}      \put(726,484){\makebox(0,0){\strut{} 0}}      \put(1286,484){\makebox(0,0){\strut{} 0.1}}      \put(1845,484){\makebox(0,0){\strut{} 0.2}}      \put(2405,484){\makebox(0,0){\strut{} 0.3}}      \put(2964,484){\makebox(0,0){\strut{} 0.4}}      \put(3524,484){\makebox(0,0){\strut{} 0.5}}      \put(4083,484){\makebox(0,0){\strut{} 0.6}}      \put(4643,484){\makebox(0,0){\strut{} 0.7}}      \put(220,2203){\rotatebox{-270}{\makebox(0,0){\strut{}Percentage of relative error}}}      \put(1628,154){\makebox(0,0){\large $\frac{2a}{\lambda_0}$}\hspace{0.4cm}  (diameter over wavelength)}    }    \gplgaddtomacro\gplfronttext{      \csname LTb\endcsname      \put(2293,2375){\makebox(0,0)[l]{\strut{}$a$}}      \put(1845,1861){\makebox(0,0)[l]{\strut{}$\Jomegar$}}      \put(2125,1475){\makebox(0,0){\strut{}Current loop}}    }    \gplbacktext
    \put(0,0){\includegraphics{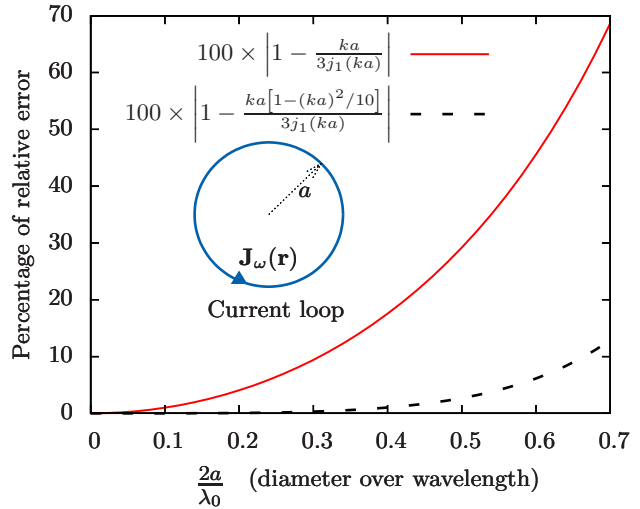}}    \gplfronttext
  \end{picture}\endgroup
 \caption{\label{fig:relerror}Relative error in the magnetic dipole moment of an infinitesimally thin circular current loop of radius $a$ (shown in the inset) due to the small $2\pi a/\lambda_0$ approximation. Solid red line: Error due to taking only the first term in the small argument expansion of the spherical Bessel function in Eq. (\ref{eq:bexact}). Such first order gives the typical integral for the magnetic dipole moment of electromagnetically small sources [see Eq. (\ref{eq:bapprox})]. Dashed black line: Error due to taking the first two terms in the expansion, i.e. Eq. (\ref{eq:bapprox}) plus Eq. (\ref{eq:k3corr}).} 
\end{center}
\end{figure}

Let us consider an infinitesimally thin circular loop of current with implicit time dependence $\exp\left(-i\omega t\right)$. The loop has radius $a$ and lies on the plane perpendicular to the $\zhat$ axis (see the inset in Fig. \ref{fig:relerror}). The expression for its current in spherical coordinates is
\begin{equation}
	\Jomegar=\phihat I_0 \delta(r-a)\frac{1}{r}\delta(\theta-\frac{\pi}{2}),
\end{equation}
where $\phihat=[-\sin\phi,\cos\phi,0]^T$, $\phi=\arctan(\frac{y}{x})$ and $\theta=\arccos(\frac{z}{r})$.

The exact value of its magnetic dipole moment is obtained after calculating the integral in Eq. (\ref{eq:bexact}): 
\begin{equation}
\label{eq:mloop}
		\mathbf{m}=\zhat\sqrt{3}I_0aj_1(ka).
\end{equation}

We obtain a first small source approximation by using Eq. (\ref{eq:bapprox}) and a more accurate second one using the incremental correction in Eq. (\ref{eq:k3corr})
\begin{equation}
	\begin{split}
		\mathbf{m}^{(1)}_{ka\ll 1}&=\zhat\sqrt{3}I_0\frac{ka^2}{3},\\
		\mathbf{m}^{(2)}_{ka\ll 1}&=\zhat\sqrt{3}I_0\frac{ka^2}{3}\left[1-(ka)^2/10\right].
	\end{split}
\end{equation}
The same results are obtained by taking terms up to $ka$ and $(ka)^3$, respectively, in the Taylor series of $j_1(ka)$ in Eq. (\ref{eq:mloop}). This latter approach relies on the existence of an exact closed form solution and is hence not general.

The exact value of Eq. (\ref{eq:mloop}) coincides with the one calculated in \cite[\S 13.3 p1881]{Morse1953} up to a numerical factor that can be traced back to a different normalization. In this simple example, the relative error incurred due to the small source approximations is equal to the relative error incurred when approximating the first order spherical Bessel function. Figure \ref{fig:relerror} shows the relative errors incurred when taking only the first term in the expansion $\left[j_1(ka)\approx ka/3\right]$ and when taking the first two terms $\left\{j_1(ka)\approx (ka/3)\times\left[1-(ka)^2/10\right]\right\}$. We see that, if we take only one term, a 10\% relative error is incurred when the diameter of the loop is approximately 30\% of the wavelength. When taking two terms, the 10\% relative error is reached when the diameter is approximately 70\% of the wavelength. We note that in this example the current is concentrated in the most exterior region of the object. When this is not the case, e.g. in a homogeneous current distribution within a sphere of diameter $2a$, the relative errors should be smaller.

\section{Results for helicity multipoles}
There is some recent interest in the use of helicity for the study of interactions between matter and electromagnetic fields \cite{Schmidt2015,FerCor2015,Cameron2014,Tischler2014,Bliokh2011b}. Due to its fundamental relationship with electromagnetic duality, the helicity formalism is also very useful when discussing dual symmetric systems \cite{FerCor2012p,FerCorTHESIS}, e.g. Huygens surfaces \cite{Pfeiffer2013,Decker2015}. We now extend our results to the dipoles of well defined helicity.

Multipoles of well defined helicity are an alternative to the multipoles of well defined parity. The two sets are related by a change of basis, which we write for both the $\qjm$ coefficients and the $\Qjm(\phat)$ functions:

\begin{widetext}
{
\begin{equation}
	\label{eq:partohel}
	\begin{split}
		\qpjm=\frac{\bjm+\ajm}{\sqrt{2}}\ &\iff\ \Qpjm(\phat)=\frac{\Xjm(\phat)+\Zjm(\phat)}{\sqrt{2}}=\frac{\mathbf{1}+i\phat\times}{\sqrt{2}}\frac{\mathbf{L}Y_{jm}}{\sqrt{j(j+1)}},\\
		\qmjm=\frac{\bjm-\ajm}{\sqrt{2}}\ &\iff\ \Qmjm(\phat)=\frac{\Xjm(\phat)-\Zjm(\phat)}{\sqrt{2}}=\frac{\mathbf{1}-i\phat\times}{\sqrt{2}}\frac{\mathbf{L}Y_{jm}}{\sqrt{j(j+1)}},\\
		\qzjm=\cjm\ &\iff \ \Qzjm(\phat)=\Wjm(\phat)=\phat Y_{jm},
	\end{split}
\end{equation}
}
\end{widetext}
where $\mathbf{1}$ is the 3$\times$3 unit matrix.

The $\mathbf{G}^{\lambda}_{jm}(\phat)$ in Eq. (\ref{eq:partohel}) and the $\mathbf{Q}_{jm}(\phat)$ have the same properties under rotations. They differ in their parity and polarization properties. Instead of eigenstates of parity, the $\mathbf{G}^{\lambda}_{jm}(\phat)$ are eigenstates of the helicity operator with eigenvalue $\lambda$. This is obvious from the rightmost expressions in Eq. (\ref{eq:partohel}) since the helicity operator $\Lambda$ in the momentum representation is $i\phat\times$:
\begin{equation}
	\Lambda=\frac{\mathbf{J}\cdot\mathbf{P}}{|\mathbf{P}|} \rightarrow i\phat\times,
\end{equation}
where $\mathbf{J}$ and ${\mathbf{P}}$ are the angular and linear momentum vector operators, respectively.

The two transverse families of this alternative basis, $\mathbf{G}^{\pm}_{jm}(\phat)$, correspond to multipolar components $g_{jm\pm}^\omega$ that radiate fields of definite polarization handedness (helicity) $\lambda=\pm1$ \cite[App. A]{FerCor2012b}. 

The extension of our dipolar results to the helicity basis is straightforward. According to the third line of Eq. (\ref{eq:partohel}), the result for $\lambda=0$ is Eq. (\ref{eq:cexact}). The exact expressions for the transverse dipoles with helicity $\lambda=\pm1$ can be obtained using Eq. (\ref{eq:bexact}), Eq. (\ref{eq:aexact}) and Eq. (\ref{eq:partohel}):

\begin{eqnarray}
\label{eq:helexact}
&&-2\pi\sqrt{6}\begin{bmatrix}g^\omega_{1\lambda}\\g^\omega_{0\lambda}\\g^\omega_{{-1}\lambda}\end{bmatrix}=\\
&&\int d^3\rr \left\{ 3j_1(kr)\rhat\times +\lambda\left[2j_0(kr)\mathbf{1}+(3\rhat\rhat^\dagger -\mathbf{1})j_2(kr)\right]\right\}\Jomegar\nonumber.
\end{eqnarray}

The approximated expressions up to order $k^2$ are:
\begin{equation}
\label{eq:helapprox}
	\begin{split}
&-2\pi\sqrt{6}\begin{bmatrix}g^\omega_{1\lambda}\\g^\omega_{0\lambda}\\g^\omega_{{-1}\lambda}\end{bmatrix}=\\
&\int d^3\rr \left\{ k\rr\times +\lambda\left[2\mathbf{1}+\frac{k^2}{5}\left(\rr\rr^\dagger -2r^2\mathbf{1}\right)\right]\right\}\Jomegar.
	\end{split}
\end{equation}

\section{Conclusion and future work}
In conclusion, we have obtained new exact expressions for the dipolar moments of a localized source distribution. These expressions are simpler than the ones reported to date. They are only marginally more complex than the typical integrals for the dipole moments of electromagnetically small sources and allow to easily obtain approximate expressions with increasing accuracy. Our results can be applied in the many areas where the dipole moments of electrical current sources are used.

In future work, we aim to obtain new exact expressions for general $j$-polar order and use them in applications like for instance in the study of the scattering properties of nanostructures.
 \section*{Acknowledgments}
I.F.-C. thanks Ms. Magda Felo for her help with the figures. S.N. acknowledges support by the Karlsruhe School of Optics \& Photonics (KSOP). We acknowledge support by Deutsche Forschungsgemeinschaft and Open Access Publishing Fund of Karlsruhe Institute of Technology. We also gratefully acknowledge financial support by the Deutsche Forschungsgemeinschaft (DFG) through CRC 1173.

\bibliographystyle{ifcbst}
\clearpage
\appendix
\section{Fields produced by time varying sources: Only Fourier components with $|\pp|=\omega/c$ contribute}\label{sec:onlyw}
We consider a electric charge and current density distributions $\rho(\rr,t)$ and $\Jrt$ embedded in an isotropic and homogeneous medium with constant and real permittivity $\epsilon$ and permeability $\mu$. We assume them to be confined in space so that $\rho(\rr,t)=0$ and $\Jrt=0$ for $|\rr|>R$. We consider the following Fourier decomposition:
\begin{equation}
	\label{eq:ff}
	\begin{split}
		&\rho(\rr,t)=\mathcal{R}\left[\int_{0^{+}}^{\infty}\frac{d\omega}{\sqrt{2\pi}}\exp\left(-i\omega t\right)\rho_\omega(\rr)\right]\\
		   &=\mathcal{R}\left[\int_{0^{+}}^{\infty}\frac{d\omega}{\sqrt{2\pi}}\exp\left(-i\omega t\right)\int \dpp \rho_\omega(\pp)\exp\left(i\pp\cdot\rr\right)\right],\\
\
		&\Jrt=\mathcal{R}\left[\int_{0^{+}}^{\infty}\frac{d\omega}{\sqrt{2\pi}}\exp\left(-i\omega t\right)\mathbf{J}_\omega(\rr)\right]\\
		   &=\mathcal{R}\left[\int_{0^{+}}^{\infty}\frac{d\omega}{\sqrt{2\pi}}\exp\left(-i\omega t\right)\int \dpp \Jomegagen\exp\left(i\pp\cdot\rr\right)\right].
	\end{split}
\end{equation}
The lower limit of the integral in $d\omega$ excludes the static case $\omega=0$.

Devaney and Wolf \cite{Devaney1974} proved that, outside the source region, the transverse parts of the electromagnetic field produced by the source at frequency $\omega$ are determined by the transverse components of $\Jomegagen$ that meet $|\pp|=\omega/c$, where $c=1/\sqrt{\epsilon\mu}$ is the speed of light in the medium.

We now provide a different proof which uses the potentials instead of the fields and shows the selection of the $|\pp|=\omega/c$ components through the appearance of a radial delta distribution. We prove that the only parts of the sources that contribute to the scalar and vector potentials in the Lorenz gauge are those in the domain $|\pp|=\omega/c$. The electric and magnetic fields obtained from the potentials are hence also determined by the components in the momentum space shell with radius $|\pp|=\omega/c$, which means that the result is independent of the choice of gauge.

In the Lorenz gauge, and with implicit monochromatic $\exp\left(-i\omega t\right)$ dependence, the sources in Eq. (\ref{eq:ff}) generate the following scalar and vector potentials: 
\begin{equation}
	\label{eq:first}
	\begin{split}
		\phi_\omega(\rr)&=\frac{1}{\epsilon}\int d^3\rr' \rho_\omega(\rr')\frac{\exp\left({ik|\rr-\rr'|}\right)}{4\pi|\rr-\rr'|}\\
		\Aomegar&=\mu\int d^3\rr'\ \Jomegarprimeall \frac{\exp\left({ik|\rr-\rr'|}\right)}{4\pi|\rr-\rr'|},
	\end{split}
\end{equation}
where $k=\omega/c$. 

Following Jackson's steps, we use the expansion of $\exp\left(ik|\rr-\rr'|\right)/(4\pi|\rr-\rr'|)$ in \cite[Eq. 9.98]{Jackson1998} 

\begin{equation}
		\frac{\exp\left(ik|\rr-\rr'|\right)}{4\pi|\rr-\rr'|}=
		ik\sum_{l=0}^\infty h^{(1)}_l(kr)j_l(kr')\sum_{m=-l}^{m=l}Y_{lm}(\rhat)Y^*_{lm}(\rhat')
\end{equation}
to get to \cite[Eq. 9.11]{Jackson1998}:
\begin{equation}
	\label{eq:A}
	\begin{split}
	&\Aomegar=\\
	&i\mu k \sum_{l,m} h^{(1)}_l(kr)Y_{lm}(\rhat)\underbrace{\intdrprime \Jomegarprime j_l(kr')Y^*_{lm}(\rhatprime)}_{\boldsymbol{\Gamma}_{lm}},
	\end{split}
\end{equation}
where $l$ and $m$ are integers, $h^{(1)}_l(\cdot)$ and $j_l(\cdot)$ are the $l$-th order spherical Hankel and Bessel functions, respectively, $\rhat=\rr/|\rr|$, $\rhatprime=\rr'/|\rr'|$, $r=|\rr|$, $r'=|\rr'|$, and $Y_{jq}$ are the scalar spherical harmonics.

Let us now consider the integral labeled as $\boldsymbol{\Gamma}_{lm}$ in Eq. (\ref{eq:A}) for a given term $(l,m)$. We use the inverse Fourier transform of $\Jomegarprimeall$
\begin{equation}
	\label{eq:invf}
	\Jomegarprimeall = \int \dpp \mathbf{J}_\omega(\pp) \exp\left(i\pp\cdot\rr'\right),
\end{equation}
and the expansion of the exponential $\exp\left(i\pp\cdot\rr'\right)$ in spherical harmonics,
\begin{equation}
	\exp\left(i\pp\cdot\rr'\right)=(4\pi)\sum_{\lbar=0}^\infty\sum_{\mbar=-\lbar}^{\mbar=\lbar}i^{\lbar} Y_{\lbar\mbar}(\rhat')Y^*_{\lbar\mbar}(\phat)j_{\lbar}(|\pp||\rr'|),
\end{equation}
to get
{\small
\begin{equation}
	\label{eq:am}
	\begin{split}
		&\frac{\boldsymbol{\Gamma}_{lm}}{4\pi}=\\
	&\sum_{\lbar\mbar}i^{\lbar}\int d^3\rr' \int \dpp \mathbf{J}_\omega(\pp) j_{\lbar}(|\pp|r')j_{l}(kr')Y_{\lbar\mbar}^*(\phat)Y_{\lbar\mbar}(\rhat')Y^*_{lm}(\rhat').
	\end{split}
\end{equation}
}

We stress that $\mathbf{J}_\omega(\pp)$ in Eq. (\ref{eq:invf}), and hence $\Jomegarprimeall$ in Eq. (\ref{eq:A}) and $\boldsymbol{\Gamma}_{lm}$ in Eq. (\ref{eq:am}), may contain contributions from momenta $\pp$ such that $|\pp|\neq \omega/c$. The following steps show that these contributions are filtered out and that $\Aomegar$ depends only on the components of $\mathbf{J}_\omega(\pp)$ with $|\pp|=\omega/c$. 

We take Eq. (\ref{eq:am}), split the integral in $d^3\rr'$ into radial and angular parts $\left(\intdrprime=\int_0^{\infty}dr'\ r'^2\int d\rhat'\right)$, and solve the angular part through the orthonormality of the spherical harmonics $\int d\rhat'Y_{\lbar\mbar}(\rhat)Y^*_{lm}(\rhat)=\delta_{\lbar l}\delta_{\mbar m}$. After this, the only term in the sum on $\lbar$ and $\mbar$ that does not vanish is the one meeting $\lbar=l$ and $\mbar=m$: 

\begin{equation}
	\label{eq:pp}
	\frac{\boldsymbol{\Gamma}_{lm}}{4\pi}=i^l \int \dpp \mathbf{J}_\omega(\pp) Y_{lm}^*(\phat) \shadedbox{\int dr'(r')^2  j_l(|\pp|r')j_l(kr')}.
\end{equation}

The crucial step is that the integral in the shaded box of Eq. (\ref{eq:pp}) has a formal solution as a radial Dirac delta distribution \cite[Eq. 4.1]{Mehrem1991}:
\begin{equation}
	\label{eq:delta}
	\int dr'(r')^2  j_l(|\pp|r')j_l(kr') =\frac{\pi}{2k^2}\delta(|\pp|-k).
\end{equation}

This $\delta(k-|\pp|)$ term discards all momenta contributions from outside the spherical shell $|\pp|=k=\omega/c$ in Eq. (\ref{eq:pp}). To show it explicitly, we split the integral in $d^3\pp$ into radial ($p=|\pp|$) and angular parts $\left(\int d^3\pp=\int_0^{\infty}dp\ p^2\int d\phat\right)$:
\begin{equation}
	\begin{split}
		\boldsymbol{\Gamma}_{lm}&=\frac{4\pi}{\sqrt{(2\pi)^3}}i^l \int d\phat \ Y_{lm}^*(\phat) \int dp\ p^2 \mathbf{J}_\omega(\pp) \frac{\pi}{2k^2}\delta(p-k)=\\
&\frac{i^l}{\sqrt{2\pi}} \int d\phat \ \mathbf{J}_\omega(\pp,|\pp|=k) Y_{lm}^*(\phat).
	\end{split}
\end{equation}

Since this conclusion holds for all values of $(l,m)$ in Eq. (\ref{eq:A}), it follows that the vector potential is completely determined by $\mathbf{J}_\omega(\pp,|\pp|=k)$, i.e., the components of $\mathbf{J}_\omega(\pp)$ on the momentum shell of radius $|\pp|=\omega/c$. 

The same conclusion is valid for the scalar potential $\phi_\omega(\rr)$ in Eq. (\ref{eq:first}). This can be seen noting that none of the steps in the previous derivation needs the fact that $\mathbf{J}_\omega(\rr)$ is a vector. The same steps can be taken for the scalar charge density $\rho_\omega(\rr)$ which generates the scalar potential in Eq. (\ref{eq:first}). Regarding its inverse Fourier transform
\begin{equation}
	\rho_\omega(\rr) = \int \dpp \rho_\omega(\pp) \exp\left(i\pp\cdot\rr\right),
\end{equation}
the conclusion in this case is that $\phi_\omega(\rr)$ only depends on the momentum components of the charge density $\rho_\omega(\pp)$ in the momentum shell of radius $|\pp|=\omega/c$. 

Since both scalar and vector potentials $(\rho_\omega(\rr),\Aomegar)$ depend only on the source Fourier components in the domain $|\pp|=\omega/c$, the same will be true for the electric and magnetic fields computed from them:
\begin{equation}
	\label{eq:EB}
	\mathbf{E}_\omega(\rr)=i\omega \mathbf{A}_\omega(\rr)-\nabla \phi_\omega (\rr),\ \mathbf{B}_\omega(\rr)=\nabla\times \mathbf{A}_\omega(\rr).
\end{equation}

It is hence clear that the conclusion is gauge independent. It is also clear that the derivation applies to both transverse and longitudinal components of the electromagnetic field, but the longitudinal electric field with $|\pp|=\omega/c$ is zero outside the source region. This can be seen in \cite[\S 13.3 p1875-1877]{Morse1953}, and in \cite[App. C]{FerCor2015c}, where the cancellation is shown to be due to the continuity equation.

\section{One term in $\bjm$, two in $\ajm$ and $\cjm$\label{sec:omte}}
We show that, for $\ajm$ and $\cjm$, only terms  with $\lbar=j-1$ or $\lbar=j+1$ can be different from zero in Eq. (\ref{eq:exact}), and that for $\bjm$, only $\lbar=j$ contributes. For this, we will write the momentum space integrals in the shaded area of Eq. (\ref{eq:exact}) as integrals of triple products of spherical harmonics. These integrals have an exact expression involving a product of two 3j-Wigner symbols. The requirement that one of the 3j-Wigner symbols be non-null results in the aforementioned relationships between $j$ and $\lbar$. 

Particularizing the shaded integrals in Eq. (\ref{eq:exact}) to $\Wjm(\phat)$ and $\Zjm(\phat)$
\begin{equation}
	\begin{split}
\label{eq:bbb}
		\intdOmegap \Wjm(\phat)^{\dagger}Y_{\lbar\mbar}(\phat)&=\intdOmegap \left[\phat Y_{jm}(\phat)\right]^{\dagger}Y_{\lbar\mbar}(\phat),\\
		\intdOmegap \Zjm^{\dagger}(\phat)Y_{\lbar\mbar}(\phat)&=\intdOmegap \left[i\phat\times\Xjm(\phat)\right]^{\dagger}Y_{\lbar\mbar}(\phat),\\
\end{split}
\end{equation}
we find that each of their three components contains either one [in the $\Wjm(\phat)$ case] or a sum of two [in the $\Zjm(\phat)$ case] triple products of spherical harmonics like $Y^*_{1p}Y^*_{jq}Y_{\lbar r}$, which can be also written\footnote{Using that $Y_{lm}^*=(-1)^mY_{l-m}$.} as $(-1)^{p+q}Y_{1-p}Y_{j-q}Y_{\lbar r}$. The result of the integral of the product of three spherical harmonics is \cite[p. 700]{Arfken1985} 
\begin{equation}
	\label{eq:triple}
	\begin{split}
	&\int d\Omega \ Y_{l_1 m_1}(\Omega)Y_{l_2 m_2}(\Omega)Y_{l_3 m_3}(\Omega)=\\&\sqrt{\frac{(2l_1+1)(2l_2+1)(2l_3+1)}{4\pi}}\begin{pmatrix}l_1&l_2&l_3\\0&0&0\end{pmatrix}\begin{pmatrix}l_1&l_2&l_3\\m_1&m_2&m_3\end{pmatrix},
	\end{split}
\end{equation}
where $\begin{pmatrix}j_1&j_2&j_3\\q_1&q_2&q_3\end{pmatrix}$ is the 3j-Wigner symbol.

In our case, we see from the right hand sides of Eq. (\ref{eq:bbb}) that $j_1=1$ from $\phat$ [see Eq. (\ref{eq:phat})], $j_2=j$ from $\Zjm(\phat)$ or $\Wjm(\phat)$, and $j_3=\lbar$ from $Y_{\lbar\mbar}(\phat)$. We now consider some of the conditions for the first 3j-symbol in Eq. (\ref{eq:triple})
\begin{equation}
	\label{eq:3j}
	\begin{pmatrix}1&j&\lbar\\0&0&0\end{pmatrix}
\end{equation}
to be different than zero. Namely \cite[p. 1056]{Messiah1958}:
\begin{equation}
	\label{eq:rest}
	|j_1-j_2|\le j_3\le j_1+j_2\implies|1-j|\le \lbar \le j+1
\end{equation}
which, if $j>0$ restricts $\lbar$ to be $j-1$, $j$ or $j+1$ and, when $j=0$ in the longitudinal case, forces $\lbar=1$. Furthermore, because of the zeros in Eq. (\ref{eq:3j}), $1+j+\lbar$ must be an integer multiple of 2, which then forbids $\lbar=j$ when $j>0$. All together we obtain for the $\Wjm(\phat)$ and $\Zjm(\phat)$ cases the restrictions:

\begin{equation}
	\label{eq:lj}
\begin{split}
	\lbar=j-1 \textrm{ or }j+1 \quad &\textrm{if $j>0$,}\\
	\lbar=1  \quad &\textrm{if $j=0$.}
  \end{split} 
 \end{equation}

In the $\Xjm(\phat)$ case, the integrals in the three components of
\begin{equation}
		\intdOmegap \Xjm^{\dagger}(\phat)Y_{\lbar\mbar}(\phat)
\end{equation}
contain a product of two spherical harmonics, but the third one can always be assumed to be the constant $1=\sqrt{4\pi}Y_{00}$. In this case the restriction of Eq. (\ref{eq:rest}) forces $\lbar=j$.
\begin{equation}
	\label{eq:lj2}
	|0-j|\le \lbar \le j+0 \implies \lbar=j.
\end{equation}

\section{Auxiliary expressions in the spherical vector basis}\label{sec:supporting}
We write a vector $\avec$ in the spherical vector basis as:
\begin{equation}
	\avec=a_{1}\ehatp+a_0\ehatz+a_{-1}\ehatm,
\end{equation}
with
\begin{equation}
	\label{eq:bvec}
	\begin{split}
	\ehatp&=-\frac{\xhat+i\yhat}{\sqrt{2}}\\
	\ehatz&=\zhat\\
	\ehatm&=\frac{\xhat-i\yhat}{\sqrt{2}}.\\
	\end{split}
\end{equation}
This choice of basis induces the following relationships between the Cartesian and spherical coordinates of $\avec$ in the spherical and Cartesian basis:
\begin{equation}
	\label{eq:cartsph}
\begin{bmatrix}a_{1}\\a_0\\a_{-1}\end{bmatrix}=
	\begin{bmatrix}
		\frac{-1}{\sqrt{2}}&\frac{i}{\sqrt{2}}&0\\
		0&0&1\\
		\frac{1}{\sqrt{2}}&\frac{i}{\sqrt{2}}&0\\
	\end{bmatrix}
\begin{bmatrix}a_x\\a_y\\a_z\end{bmatrix},
\begin{bmatrix}a_x\\a_y\\a_z\end{bmatrix}=
	\begin{bmatrix}
		\frac{-1}{\sqrt{2}}&0&\frac{1}{\sqrt{2}}\\
		\frac{-i}{\sqrt{2}}&0&\frac{-i}{\sqrt{2}}\\
		0&1&0\\
	\end{bmatrix}
\begin{bmatrix}a_{1}\\a_0\\a_{-1}\end{bmatrix}.
\end{equation}
In the spherical basis, the components of the cross product of two vectors are \footnote{Equation (\ref{eq:cross}) is obtained from the relations $\ehatp\times\ehatm=i\ehatz$, $\ehatp\times\ehatz=i\ehatp$ and $\ehatm\times\ehatz=-i\ehatm$, which follow from Eq. (\ref{eq:bvec}) and the cross products between the Cartesian basis vectors $\{\xhat,\yhat,\zhat\}$. }: 
\begin{equation}
	\label{eq:cross}
\avec\times\bvec=i\begin{bmatrix}a_{1}b_0-a_0b_{1}\\a_{1}b_{-1}-a_{-1}b_{1}\\a_0b_{-1}-a_{-1}b_0\end{bmatrix}.
\end{equation}

Let us now write some explicit expressions for $\phat$ and $\Xjm(\phat)$ that we use in the text.
\begin{equation}
	\label{eq:phat}
\phat=\frac{\pp}{|\pp|}=\begin{bmatrix}\hat{p}_1\\\hat{p}_0\\\hat{p}_{-1}\end{bmatrix}=2\sqrt{\frac{\pi}{3}}\begin{bmatrix}-Y_{1-1}\\Y_{10}\\-Y_{11}\end{bmatrix}=2\sqrt{\frac{\pi}{3}}\begin{bmatrix}Y_{11}^*\\Y_{10}^*\\Y_{1-1}^*\end{bmatrix},
\end{equation}

\begin{equation}
	\label{eq:xjm}
\Xjm(\phat)=\frac{1}{\sqrt{j(j+1)}}\begin{bmatrix}-\sqrt{\frac{j(j+1)-m(m-1)}{2}}Y_{j(m-1)}(\phat)\\ \\mY_{jm}(\phat)\\\\\sqrt{\frac{j(j+1)-m(m+1)}{2}}Y_{j(m+1)}(\phat)\end{bmatrix}.
\end{equation}

Equation (\ref{eq:phat}) follows Eq. (\ref{eq:cartsph}), the expressions of $Y_{1m}$ in Cartesian coordinates and the property $Y_{lq}^*=(-1)^qY_{l-q}$. Equation (\ref{eq:xjm}) follows from the definition of $\Xjm$ in Eq. (\ref{eq:xzw}) and the expression of the angular momentum vector operator $\mathbf{L}$ in spherical coordinates
\begin{equation}
\mathbf{L}=\begin{bmatrix}\frac{-L_x+iL_y}{\sqrt{2}}\\ \\L_z\\ \\\frac{L_x+iL_y}{\sqrt{2}}\end{bmatrix}
=\begin{bmatrix}-\frac{L_{\mathrm{down}}}{\sqrt{2}}\\ \\L_0\\ \\\frac{L_{\mathrm{up}}}{\sqrt{2}}\end{bmatrix},
\end{equation}
where $L_{\mathrm{up}}=L_x+iL_y$ and $L_{\mathrm{down}}=L_x-iL_y$ are the angular momentum ladder operators
{\small
\begin{equation}
	\begin{split}
		L_{\mathrm{up}}Y_{jm} &=\left\{
			\begin{array}{cc}
				\sqrt{j(j+1)-m(m+1)}Y_{j(m+1)}&\quad \text{if } |m+1|\le j\\
				0&\quad \mathrm{else}\\
			\end{array}\right.,\\
		L_{\mathrm{down}}Y_{jm} &=\left\{
			\begin{array}{cc}
				\sqrt{j(j+1)-m(m-1)}Y_{j(m-1)}&\quad \text{if } |m-1|\le j\\
				0&\quad \mathrm{else}\\
			\end{array}\right..\\
	\end{split}
\end{equation}
}

\section{Expression of selected $\qjm$ tensors as spatial integrals\label{sec:appB}}
\subsection{Case $a_{1m}$}\label{sec:a1m}
As shown in App. \ref{sec:omte} only $\lbar=0$ and $\lbar=2$ can have non zero contributions to $a_{1m}$. That is 
\begin{equation}
		a_{1m}^\omega={a_{1m}^{\omega}}^{\lbar=0}+{a_{1m}^{\omega}}^{\lbar=2}.
\end{equation}

We start with $\lbar=0$. From Eq. (\ref{eq:exact}), and since $Y_{00}=1/\sqrt{4\pi}$:
\begin{equation}
	\label{eq:a1}
	a_{1m}^{\lbar=0}=\factorfourier\shadedbox{\intdOmegap \mathbf{Z}_{1m}^{\dagger}(\phat)}\intdr \Jomegar j_{0}(kr).
\end{equation}

The explicit expressions for $\mathbf{Z}_{1m}=i\phat\times\mathbf{X}_{1m}$ are
\begin{equation}
	 \begin{split}
	 &m=1\rightarrow\ i\phat\times \mathbf{X}_{11}(\phat)=-\sqrt{\frac{2\pi}{3}}\begin{bmatrix}Y_{10}^2-Y_{11}Y_{1-1}\\-Y_{11}Y_{10}\\Y_{11}^2\end{bmatrix},\\
		 &m=0\rightarrow\ i\phat\times \mathbf{X}_{10}(\phat)=-\sqrt{\frac{2\pi}{3}}\begin{bmatrix}Y_{10}Y_{1-1}\\-2Y_{11}Y_{1-1}\\Y_{10}Y_{11}\end{bmatrix},\\
		 &m=-1\rightarrow\ i\phat\times \mathbf{X}_{1-1}(\phat)=-\sqrt{\frac{2\pi}{3}}\begin{bmatrix}Y_{1-1}^2\\-Y_{10}Y_{1-1}\\Y_{10}^2-Y_{11}Y_{1-1}\end{bmatrix}.\\
	 \end{split}
 \end{equation}
The relationship $Y^*_{lq}=(-1)^qY_{l-q}$, and the orthonormality of the spherical harmonics allow us to solve the momentum space integrals in the shaded area of Eq. (\ref{eq:a1}), and immediately reach
 \begin{equation}
	 {\begin{bmatrix}
	 \aoo\\\aoz\\\aom\end{bmatrix}}^{\lbar=0}=-\frac{1}{\pi\sqrt{3}}\intdr \Jomegar j_0(kr).
\end{equation}

In the $\lbar=2$ case
\begin{equation}
	\label{eq:a12}
	\begin{split}
	&\frac{\sqrt{(2\pi)^3}}{4\pi}a_{1m}^{\lbar=2}=\\&
		-\sum_{\mbar=-2}^{\mbar=2}\shadedbox{\intdOmegap\mathbf{Z}_{1m}^{\dagger}(\phat)Y_{2\mbar}}\intdr\Jomegar Y^*_{2\mbar}j_2(kr),
	\end{split}
\end{equation}
the shaded momentum space integrals contain triple products of spherical harmonics and can be solved using 

\begin{equation}
	\label{eq:triple2}
	\begin{split}
	&\int d\phat Y_{l_1 m_1}(\phat)Y_{l_2 m_2}(\phat)Y_{l_3 m_3}(\phat)=\\&\sqrt{\frac{(2l_1+1)(2l_2+1)(2l_3+1)}{4\pi}}\begin{pmatrix}l_1&l_2&l_3\\0&0&0\end{pmatrix}\begin{pmatrix}l_1&l_2&l_3\\m_1&m_2&m_3\end{pmatrix},
	\end{split}
\end{equation}
where $\begin{pmatrix}j_1&j_2&j_3\\q_1&q_2&q_3\end{pmatrix}$ is the 3j-Wigner symbol.

They result in five vectors for each $m$ case, which we list here as row vectors. From top to bottom, the row vectors correspond to $\mbar=2,1,0,-1,-2$:

\begin{equation}
	\begin{split}
	m=1&\rightarrow\ \frac{-1}{\sqrt{30}}\begin{array}{ccccc} (&0&0&\sqrt{6}&)\\(&0&-\sqrt{3}&0&)\\(&1&0&0&)\\(&0&0&0&)\\(&0&0&0&)\end{array}\\
	m=0&\rightarrow\ \frac{-1}{\sqrt{30}}\begin{array}{ccccc} (&0&0&0&)\\(&0&0&\sqrt{3}&)\\(&0&-2&0&)\\(&\sqrt{3}&0&0&)\\(&0&0&0&)\end{array}\\
	m=-1&\rightarrow\ \frac{-1}{\sqrt{30}}\begin{array}{ccccc} (&0&0&0&)\\(&0&0&0&)\\(&0&0&1&)\\(&0&-\sqrt{3}&0&)\\(&\sqrt{6}&0&0&)\end{array}\\
	\end{split}
\end{equation}

The summation in $\mbar$ in Eq. (\ref{eq:a12}) can now be done. With $\Jomegar=[\Jone ,\Jzero ,\Jmone ]^T$, and $Y^*_{2m}=(-1)^mY_{2-m}$, it reads 

\begin{widetext}
\begin{equation}
\label{eq:aaa}
	\begin{split}
			{\aoo}^{\lbar=2}&=\factorfourier\frac{4\pi}{\sqrt{30}}\intdr\left(\sqrt{6}\Jmone Y_{2-2}+\sqrt{3}\Jzero Y_{2-1}+\Jone Y_{20}\right)j_2(kr),\\
			{\aoz}^{\lbar=2}&=\factorfourier\frac{4\pi}{\sqrt{30}}\intdr\left(-\sqrt{3}\Jmone Y_{2-1}-2\Jzero Y_{20}-\sqrt{3}\Jone Y_{21}\right)j_2(kr),\\
			{\aom}^{\lbar=2}&=\factorfourier\frac{4\pi}{\sqrt{30}}\intdr\left(\Jmone Y_{20}+\sqrt{3}\Jzero Y_{21}+\sqrt{6}\Jone Y_{22}\right)j_2(kr).\\
	\end{split}
\end{equation}

\end{widetext}

We now use the following relationships: 
\begin{equation}
	\begin{split}
		Y_{22}=\sqrt{\frac{10\pi}{3}}Y_{11}^2&,\ Y_{21}=\sqrt{\frac{20\pi}{3}}Y_{10}Y_{11}\\
		Y_{20}&=\sqrt{5\pi}\left(Y_{10}^2-\frac{1}{4\pi}\right)\\
		Y_{2-2}=\sqrt{\frac{10\pi}{3}}Y_{1-1}^2&,\ Y_{2-1}=\sqrt{\frac{20\pi}{3}}Y_{10}Y_{1-1},
	\end{split}
\end{equation}
which we substitute in Eq. (\ref{eq:aaa}) and get
\begin{widetext}
\vspace{0.5cm}
\begin{equation}
	\label{eq:almost}
	\begin{split}
		{\aoo}^{\lbar=2}&=\frac{2}{\sqrt{3}}\intdr\left[Y_{1-1}\shadedbox{\left(\Jmone Y_{1-1}+\Jzero Y_{10}\right)}+\frac{\Jone }{2}\left(Y_{10}^2-\frac{1}{4\pi}\right)\right]j_2(kr),\\
			{\aoz}^{\lbar=2}&=\frac{-2}{\sqrt{3}}\intdr\left[Y_{10}\shadedbox{\left(\Jmone Y_{1-1}+\Jone Y_{11}\right)}+\Jzero \left(Y_{10}^2-\frac{1}{4\pi}\right)\right]j_2(kr),\\
			{\aom}^{\lbar=2}&=\frac{2}{\sqrt{3}}\intdr\left[Y_{11}\shadedbox{\left(\Jzero Y_{10}+\Jone Y_{11}\right)}+\frac{\Jmone }{2}\left(Y_{10}^2-\frac{1}{4\pi}\right)\right]j_2(kr).\\
	\end{split}
\end{equation}
\end{widetext}

The expressions in the shaded areas of Eq. (\ref{eq:almost}) can be completed to $Y_{11}\Jone +Y_{10}\Jzero +Y_{1-1}\Jmone $ using terms to their right. In the case of the ${\aoz}^{\lbar=2}$ the completion is straightforward. For the other two cases one uses that
\begin{equation}
	\frac{3}{4\pi}=|Y_{10}|^2+|Y_{11}|^2+|Y_{1-1}|^2\implies Y_{10}^2-\frac{1}{4\pi}=\frac{1}{2\pi}+2Y_{11}Y_{1-1}.
	\end{equation}

Finally, noting that 
\begin{equation}
	Y_{11}\Jone +Y_{10}\Jzero +Y_{1-1}\Jmone =\frac{1}{2}\sqrt{\frac{3}{\pi}}\left[\rhat^{\dagger}\Jomegar\right],
\end{equation}
we reach the final result
\begin{equation}
	{\begin{bmatrix}
	\aoo\\\aoz\\\aom\end{bmatrix}}^{\lbar=2}=-\frac{1}{2\pi\sqrt{3}}\intdr\left\{3\left[\rhat^{\dagger}\Jomegar\right]\rhat-\Jomegar\right\}j_2(kr).\\
\end{equation}

The sum of the two contributions can be manipulated with the aid of the recursion relations between spherical Bessel functions:
\begin{equation}
	\begin{split}
		\frac{2l+1}{x}j_l(x)&=j_{l-1}(x)+j_{l+1}(x),\\
		(2l+1)\frac{d}{dx}j_l(x)&=lj_{l-1}(x)-(l+1)j_{l+1}(x),\\
	\end{split}
\end{equation}
to get, with the definitions $\Jomegarhat=\left[\rhat^{\dagger}\Jomegar\right]\rhat$ and $\Jomegart=\Jomegar-\Jomegarhat$, 
\begin{equation}
	\begin{split}
\begin{bmatrix}
\aoo\\\aoz\\\aom\end{bmatrix}&=-\frac{1}{2\pi\sqrt{3}}\intdr \Jomegarhat \frac{6}{kr}j_1(kr)\\
	 &-\frac{1}{2\pi\sqrt{3}}\intdr3\Jomegart\left(\frac{1}{kr}+\frac{d}{d(kr)}\right)j_1(kr).
	\end{split}
\end{equation}

\subsection{Case $c_{00}$}\label{eq:c00}
In the $j=0$ case the contribution corresponding to $\lbar=j-1=-1$ does not exist (see App. \ref{sec:omte}), so the only contribution comes from $\lbar=1$:
\begin{equation}
	\begin{split}
		i\sqrt{\frac{\pi}{2}}c_{00}&=\sum_{\mbar=-1}^{\mbar=1}\intdOmegap \mathbf{W}_{00}^\dagger(\phat)\intdr \Jomegar Y_{1\mbar}^*j_1(kr)\\
					   &=\sum_{\mbar=-1}^{\mbar=1}\shadedbox{\intdOmegap \left(\phat \right)^\dagger Y_{1\mbar}(\phat)} \intdr \Jomegar Y_{1\mbar}^*j_1(kr).
	\end{split}
\end{equation}

The integrals in the shaded area are conveniently solved using Eq. (\ref{eq:phat}) and the orthonormality of the spherical harmonics. After the sum in $\mbar$ we get:

\begin{equation}
	\begin{split}
		i\sqrt{\frac{\pi}{2}}c_{00}&=\intdr2\sqrt{\frac{\pi}{3}}\left(-\Jmone Y_{11}^*+\Jzero Y_{10}^*-\Jone Y_{1-1}^*\right)j_1(kr)\\&=\intdr \left[\rhat^\dagger \Jomegar\right]j_1(kr).
	\end{split}
\end{equation}

The first term in a small $kr$ expansion of $c_{00}$ will be of order $k$:
\begin{equation}
	c_{00}\approx -i\sqrt{\frac{2}{\pi}}\frac{k}{3} \intdr \left[\rr^\dagger \Jomegar\right].
\end{equation}

\subsection{Case $c_{1m}$}\label{sec:c1m}
As in Sec. \ref{sec:a1m}, we split the two contributions:
\begin{equation}
	c_{1m}^\omega={c_{1m}^{\omega}}^{\lbar=0}+{c_{1m}^{\omega}}^{\lbar=2}.
\end{equation}

For $\lbar=0$, and recalling that $Y_{00}=1/\sqrt{4\pi}$:
\begin{equation}
	\begin{split}
		c_{1m}^{\lbar=0}&=\frac{4\pi}{\sqrt{(2\pi)^3}}\intdOmegap \mathbf{W}_{1m}(\phat)^\dagger \frac{1}{\sqrt{4\pi}}\intdr \Jomegar\frac{1}{\sqrt{4\pi}}j_0(kr),\\
	&=\factorfourier\shadedbox{\intdOmegap \left[\phat Y_{1m}(\phat)\right]^\dagger}\intdr \Jomegar j_0(kr).\\
\end{split}
\end{equation}

The result of the integrals in the shaded area above is:
\begin{equation}
m=1:\ 2\sqrt{\frac{\pi}{3}}\begin{bmatrix}1\\0\\0\end{bmatrix},\
m=0:\ 2\sqrt{\frac{\pi}{3}}\begin{bmatrix}0\\1\\0\end{bmatrix},\
m=-1:\ 2\sqrt{\frac{\pi}{3}}\begin{bmatrix}0\\0\\1\end{bmatrix}.
\end{equation}

With which we reach:

\begin{equation}
{\begin{bmatrix}
\coo\\\coz\\\com\end{bmatrix}}^{\lbar=0}= \frac{1}{\pi\sqrt{6}}\intdr \Jomegar j_0(kr).
\end{equation}

For $\lbar=2$:
\begin{equation}
	\begin{split}
		&c_{1m}^{\lbar=2}=\\&
		\frac{-4\pi}{\sqrt{(2\pi)^3}}\sum_{\mbar=-2}^{\mbar=2}\shadedbox{\intdOmegap\left[\phat Y_{1m}(\phat)\right]^\dagger Y_{2\mbar}}\intdr\Jomegar Y^*_{2\mbar}j_2(kr),
	\end{split}
\end{equation}
the shaded momentum space integrals contain triple products of spherical harmonics and can be solved using Eq. (\ref{eq:triple2}). They result in five vectors for each $m$ case, which we list here as row vectors. From top to bottom, the row vectors corresponds to $\mbar=2,1,0,-1,-2$:

\begin{equation}
	\begin{split}
	m=1&\rightarrow\ \frac{-1}{\sqrt{15}}\begin{array}{ccccc} (&0&0&\sqrt{6}&)\\(&0&-\sqrt{3}&0&)\\(&1&0&0&)\\(&0&0&0&)\\(&0&0&0&)\end{array}\\
	m=0&\rightarrow\ \frac{-1}{\sqrt{15}}\begin{array}{ccccc} (&0&0&0&)\\(&0&0&\sqrt{3}&)\\(&0&-2&0&)\\(&\sqrt{3}&0&0&)\\(&0&0&0&)\end{array}\\
	m=-1&\rightarrow\ \frac{-1}{\sqrt{15}}\begin{array}{ccccc} (&0&0&0&)\\(&0&0&0&)\\(&0&0&1&)\\(&0&-\sqrt{3}&0&)\\(&\sqrt{6}&0&0&)\end{array}\\
	\end{split}
\end{equation}

The following result is reached after taking steps parallel to those taken in Sec. \ref{sec:a1m} for $a_{1m}^{\lbar=2}$:
\begin{equation}
	\label{eq:cexactpp}
		\begin{split}
	\begin{bmatrix}\coo\\\coz\\\com\end{bmatrix}&=\underbrace{\frac{1}{\pi\sqrt{6}}\intdr \Jomegar j_0(kr)}_{\lbar=0} \\
		&-\underbrace{\frac{1}{\pi\sqrt{6}}\intdr\left\{3\left[\rhat^{\dagger}\Jomegar\right]\rhat-\Jomegar\right\}j_2(kr)}_{\lbar=2}.
		\end{split}
\end{equation}
 \end{document}